%% file: multitask_w2v2_Kunesova-Zajic_ICASSP2023.tex
\documentclass{article}
\usepackage[preprint]{spconf}
\usepackage{amsmath,graphicx}
\ninept 
\usepackage{subfig} 

\usepackage{url}

\usepackage{booktabs}
\usepackage{siunitx} 
\usepackage{tabularx}

\newcolumntype{H}{>{\setbox0=\hbox\bgroup}c<{\egroup}@{}}

\usepackage{adjustbox} 

\usepackage[above, below]{placeins} 

\usepackage{comment}

\usepackage{microtype} 

\newcommand\extrafootertext[1]{%
    \bgroup
    \renewcommand\thefootnote{\fnsymbol{footnote}}%
    \renewcommand\thempfootnote{\fnsymbol{mpfootnote}}%
    \footnotetext[0]{#1}%
    \egroup
}

\title{Multitask Detection of Speaker Changes, Overlapping Speech and Voice Activity Using wav2vec 2.0} 

\name{ 
\begin{tabular}{c} 
    Marie Kune{\v s}ov{\'a} 
    and
    Zbyn{\v e}k Zaj{\'i}c%
    \thanks{This research was supported by the Czech Ministry of Interior, project ROZKAZ (VJ01010108) and by the Czech Ministry of Education, Youth and Sports, Project No. (LM2023062) LINDAT/CLARIAH-CZ.
    Computational resources were provided by the e-INFRA CZ project (ID:90140), supported by the Ministry of Education, Youth and Sports of the Czech Republic.
    }
\end{tabular}
}

\address{
New Technologies for the Information Society and Department of Cybernetics,\\
Faculty of Applied Sciences, University of West Bohemia, Pilsen, Czech Republic}

\copyrightnotice{\copyright\ 2023 IEEE} 
\toappear{To appear in {\it Proc.\ ICASSP 2023, June 04-10, 2023, Rhodes Island, Greece}}

\begin{document}
\topmargin=0mm 

\maketitle
\begin{abstract} 

Self-supervised learning approaches have lately achieved great success on a broad spectrum of machine learning problems. In the field of speech processing, one of the most successful recent self-supervised models is wav2vec 2.0. In this paper, we explore the effectiveness of this model on three basic speech classification tasks: speaker change detection, overlapped speech detection, and voice activity detection. 
First, we concentrate on only one task -- speaker change detection -- where our proposed system surpasses the previously reported results on four different corpora, and achieves comparable performance even when trained on out-of-domain data from an artificially designed dataset.
Then we expand our approach to tackle all three tasks in a single multitask system with state-of-the-art performance on the AMI corpus. 
The implementation of the algorithms in this paper is publicly available at \url{https://github.com/mkunes/w2v2_audioFrameClassification}.

\end{abstract}

\section{Introduction}

Speaker change detection (SCD), overlapping speech detection (OSD), and voice activity detection (VAD) are three basic speech processing tasks that are relevant for a variety of different speech applications. SCD is the task of finding the points in a conversation where the speaker is changing, while OSD is concerned with identifying intervals where multiple speakers are active at the same time. Both of these are particularly important to speaker diarization~\cite{ZZ_SD_Specom_2016_Springer,ICASSP20_Bullock_OSD_AMI_DH1}, as well as other tasks related to processing multi-speaker audio~\cite{ICASSP20_SCD_CNN_Broadcast}. Voice activity detection simply distinguishes between speech and non-speech and has a use in nearly all speech processing. 

In the past, approaches to SCD have mainly consisted of computing a distance between two sliding windows over the signal and then locating peaks~\cite{IS13_SD_Idiap} or detecting the differences in pitch~\cite{ICASSP19_SCD_FF}. Nowadays, the prevailing approach is to use deep learning. First attempts used precomputed features based on i-vectors or x-vectors~\cite{ICASSP20_SCD_CNN_Broadcast}, Mel-frequency cepstral coefficients (MFCCs)~\cite{ICASSP19_SCD_FF}, spectrograms~\cite{2017_ICASSP_Hruz_Zajic}, or combinations of multiple types of features~\cite{ICASSP22_SCD_AMI}, even including lexical information gained from automated transcripts~\cite{ZZ_LMscore2018_Springer,ACM21_HybridSCD}. Different neural network model architectures have been applied, such as LSTM~\cite{Specom2018_HruzHlavac_SCD_LSTM_Springer}, CNN~\cite{2017_ICASSP_Hruz_Zajic}, or sequence-level modeling methods~\cite{SPL22_SCD_AMI_DH1}.

A similar shift from conventional techniques to deep learning has also occurred for OSD. Recent approaches include
convolutional neural networks~\cite{Kunesova2019_SPECOM_Springer} 
or LSTMs~\cite{ICASSP20_Bullock_OSD_AMI_DH1}, and the input can be in the form of MFCCs~\cite{Cornell2020}, spectrograms~\cite{Kazimirova2018}, x-vectors~\cite{Mateju22_interspeech} or raw waveforms~\cite{IS21_OSD_Bredin_AMI_DH3}. 

VAD has been the target of a large amount of research for many years~\cite{Ramirez2007,ICASSP16_VAD_comparativeStudy}, but these days it is rarely the main focus by itself. Rather, it usually appears as one part or a by-product of a more complex system~\cite{Cornell2020,ICASP21_BW-EDA-EEND_online}.

\begin{figure}[t]
    \centering
    \includegraphics[width=0.85\linewidth]{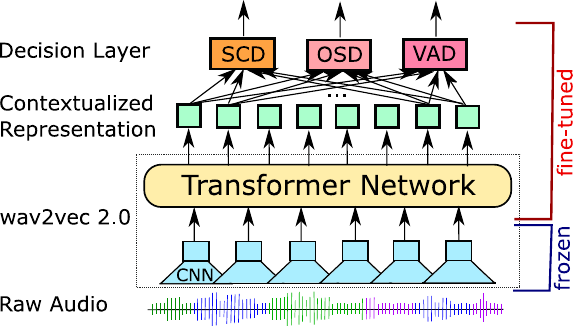}
    \caption{Illustration of the multitask wav2vec 2.0 detector of speaker changes, voice activity, and overlapping speech. The model outputs a set of labels for each audio frame (every 20\,ms).} 
    \label{fig:w2v_diagramZZ}
    \vspace{-1em} 
\end{figure}

In this paper, we first propose an end-to-end approach for SCD using the transformer network concept (which has recently seen great success on a variety of tasks, including but not limited to speech processing~\cite{Liu2021_s3plr_TERA}) -- specifically using the wav2vec 2.0~\cite{baevski2020wav2vec} framework -- and evaluate it on four commonly used conversational speech corpora. 
Then we expand the approach to also perform two other tasks (OSD and VAD) in a single multitask system, as illustrated in Figure~\ref{fig:w2v_diagramZZ}. These three tasks are closely related to each other, and it has been previously demonstrated that joint learning can improve learning efficiency and prediction accuracy compared to task-specific models~\cite{IS2022_MultiTask_ASR_SD}.

\extrafootertext{\copyright\ 2023 IEEE. Personal use of this material is permitted. Permission from IEEE must be obtained for all other uses, in any current or future media, including reprinting/republishing this material for advertising or promotional purposes, creating new collective works, for resale or redistribution to servers or lists, or reuse of any copyrighted component of this work in other works.}

\section{Single-task model for SCD}
Wav2vec 2.0 (hereafter referred to as ``wav2vec2'') is a trans\-for\-mer-based self-supervised framework for speech representation, which has been used for a wide range of speech processing tasks, such as automatic speech recognition~\cite{IS22_ASR_wav2vec2_Lehecka} and many others~\cite{IS21_s3plr_SUPERB}.

Inspired by our previous work~\cite{Kunesova2022TSD_Springer} on prosodic boundary detection, we treat the SCD problem as an audio frame classification task. We use the base model \emph{wav2vec2-base} and a larger cross-lingual (XLSR) model \emph{wav2vec2-large-xlsr-53}~\cite{conneau2020wav2vec2large}\footnote{Downloaded from \url{https://huggingface.co/facebook/wav2vec2-base} and  \url{https://huggingface.co/facebook/wav2vec2-large-xlsr-53}, respectively} with an additional last decision layer and fine-tune them for SCD using the HuggingFace Transformers~\cite{wolf2020transformers} library, in a similar manner to our aforementioned previous paper\footnote{Our code is available at \url{https://github.com/mkunes/w2v2_audioFrameClassification}.}.

A diagram of the model can be seen in Figure~\ref{fig:w2v_diagramZZ} (the figure shows the final multitask model; the single-task version has only one output but is otherwise identical). 
The wav2vec2 model creates a contextual representation of the 16kHz input signal.
The outputs from the transformer are fully connected to the decision layer (one neuron with a linear activation function), which outputs information about the speaker changes in each audio frame -- every 20\,ms, as per the pre-trained wav2vec2 model.
Due to the character of the labeling function (see below), the model is trained for regression (with mean square error loss) rather than a simple binary classification.

Because of the relatively high memory requirements of the wav2vec2 models, the input signal is given in segments of 20\,seconds, with a 10-second overlap between segments. Then when the resulting predictions are joined back together for evaluation, we use the middle part of each segment and discard the duplicate 5\,s intervals at the edges. This ensures that there is always sufficient context on both sides of a potential speaker change point.

\subsection{Reference labels for SCD}
\label{sec_annotation}

In our work, we define a \emph{speaker change point} as either the beginning or end of a speaker turn, regardless of whether any other speakers are active. In other words, we do not only seek boundaries between two different speakers but also between a speaker and silence or between one speaker and multiple.

Reference labels for the SCD task are based on the annotation files of each conversation, specifically given in the Rich Transcription Time Marked (RTTM) format: each individual record from the RTTM files is treated as a single unbroken speaker turn of the specified speaker. However, during training, we ignore short gaps (under one second) between two turns of the same speaker. We found that this helps to prevent the model from becoming too sensitive and reporting even brief pauses between words as speaker change points. 

\begin{figure}
    \centering
    \adjincludegraphics[width=0.5\textwidth,trim={{0.1\width} {0.05\height} {.05\width} {0.05\height}},clip]{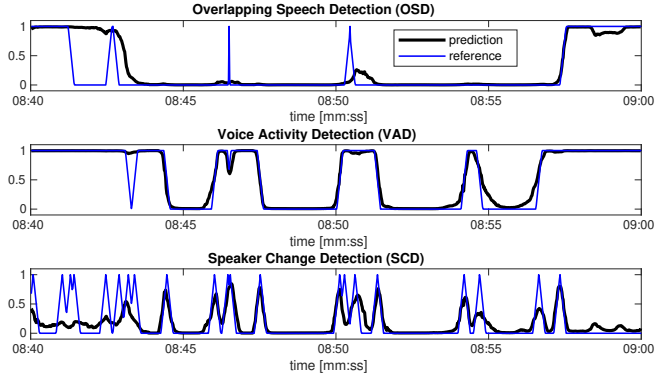}
    \caption{Output (black) and fuzzy reference (blue) of the multitask model for one audio segment of the AMI file ``EN2002b''.}
    \label{fig:multitask_example}
\end{figure}

We have also observed that the real conversational corpora that we use (described in section~\ref{sec:data}) contain time inaccuracies or inconsistencies in the reference transcripts, as human annotators sometimes lack the precision needed for training a neural network. In our previous work~\cite{2017_ICASSP_Hruz_Zajic}, we developed a fuzzy labeling strategy to cope with the uncertainty introduced by human annotators, which we also apply here:
speaker change points are given a reference label with a value of 1, which linearly decreases to zero over an interval of $\pm 0.2$\,s around each boundary. Audio frames more than 0.2\,s away from the nearest speaker change point are given a label of 0. An example of this triangular labeling can be seen in the bottom plot of Figure~\ref{fig:multitask_example}. 

During evaluation, detected speaker change points are found as peaks in the predicted labels higher than a set threshold. 
We do not perform any other post-processing of the outputs. 

\section{Datasets}
\label{sec:data}

We decided to test our proposed system on several frequently used corpora of conversational speech, where we can compare our approach with the recent results of other authors. Section \ref{sec:data-real} describes the corpora and their division into the training and testing sets. Additionally, section~\ref{subsec_artData} describes an artificial dataset which we used as an alternative option for training.

\subsection{Real conversation data}
\label{sec:data-real}

We used four English-language corpora of conversational speech:

AMI Meetings Corpus (\textbf{AMI})~\cite{LREC07_AMI_corpus} consists of 100 hours of meetings between 3--5 speakers. For our experiments, we used the ``headset mix'' recordings, as and the training and development sets followed the official ``Full-corpus partition of meetings''\footnote{\url{https://groups.inf.ed.ac.uk/ami/corpus/datasets.shtml}}. For evaluation on the AMI corpus, we used the \texttt{pyannote} library plugin \texttt{pyannote.db.odessa.ami}\footnote{\url{https://github.com/pyannote/pyannote-db-odessa-ami}} with its built-in reference annotations (the test set of which is a subset of the official one).
  
First DIHARD Challenge data (\textbf{DH-I})~\cite{DIHARD_plan,dihard_seedlings} contains audio from 12 challenging domains that range from clean (audiobooks, radio interviews) to noisy recordings (child language recordings, restaurant conversations). For our purposes, we split the original DIHARD I development data into a development and training set, using the partitioning published by~\cite{SPL22_SCD_AMI_DH1}\footnote{\url{https://github.com/zhiyunfan/SEQ- SCD/tree/master/data/dihard1}}.

Second DIHARD Challenge data (\textbf{DH-II})~\cite{DIHARD2_Ryant2019,dihard_seedlings} is a successor to DH-I and contains data from 13 audio sources. The contents are mostly identical to DH-I, but with some additions and corrections. We followed the example of \cite{ICASSP20_Bullock_OSD_AMI_DH1} and split the development data into 2/3 for training and 1/3 for development.

Finally, we also used the American English subset of the \textbf{CALLHOME}~\cite{Callhome_LDC1997} corpus -- a dataset of narrow-band telephone conversations, mostly between two speakers, which we converted to 16 kHz for this experiment. The training and test sets were the same as     in~\cite{2017_ICASSP_Hruz_Zajic,Specom2018_HruzHlavac_SCD_LSTM_Springer}: 43 and 77 files, respectively.

\subsection{Artificial dataset}
\label{subsec_artData}

To illustrate the behavior of our system on out-of-domain data, we designed a synthetic training dataset made from the LibriSpeech~\cite{LibriSpeech2015} corpus, where we can control the speaker change points and also ensure that reference labels are accurate. The creation process is similar to the creation of the artificial dataset used in~\cite{Kunesova2019_SPECOM_Springer}, but specifically tailored for speaker change detection.

\begin{figure}[t]
    \centering
    \adjincludegraphics[width=0.45\textwidth,trim={0 {0.05\height} 0 {0.03\height}},clip]{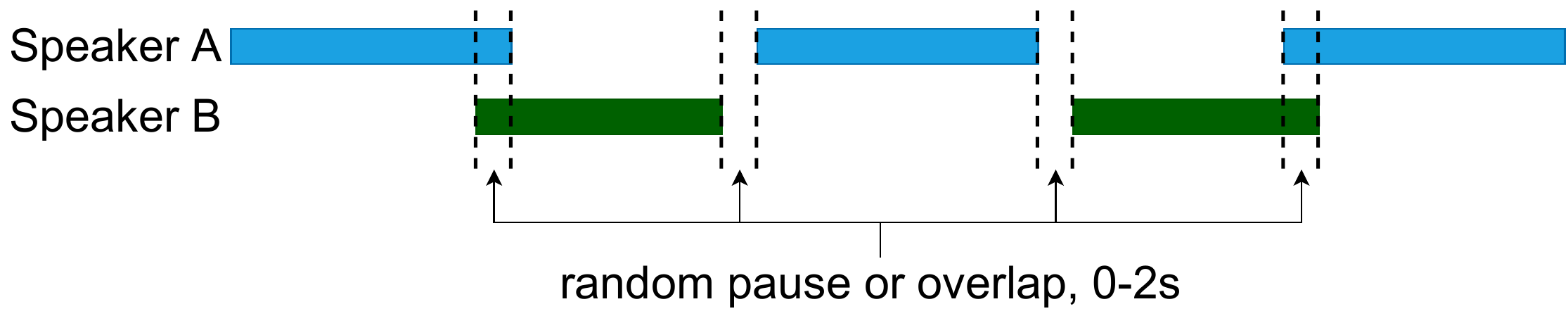}
    \caption{Artificial training data created from the LibriSpeech corpus. Each colored rectangle represents one utterance.}
    \label{fig:artificial_libri_data}
\end{figure}

The artificial dataset was created from the ``train-other-500'' subset of LibriSpeech, by always concatenating five different utterances (individual audio files, typically around 5--15 seconds long) from two different speakers in an A-B-A-B-A pattern (see Fig.~\ref{fig:artificial_libri_data}). Between each pair of utterances in the sequence, we either insert a random pause of up to 2 seconds or partially overlap them with an overlap of up to 2 seconds. This is meant to simulate a relatively realistic amount of pauses and overlaps in a real conversation. We used 500 such sequences in our training, with a total duration of approximately 8 hours.

The duration of the pauses or overlaps is uniformly distributed and the leading and trailing silence at the start and end of each utterance was linearly tapered to avoid discernible seams. Unlike \cite{Kunesova2019_SPECOM_Springer}, we did not apply additional noise or reverberation, as we did not find it to be very beneficial in this case.

\section{Results of the single-task SCD approach}

\input{table_SCD_results} 

\input{table_SCD_others_results} 

Predicted speaker change points were evaluated in terms of audio segmentation, as segment purity (Pur), coverage (Cov), and F1, using the Python library \texttt{pyannote.metrics}\footnote{\label{note_Pyannote}Downloaded from: \url{https://pyannote.github.io/}}~\cite{IS17_Bredin_Pyannote.metrics}. 
Purity measures how homogeneous the segments are, and coverage expresses whether each speaker turn is fully contained within one segment. F1 is the harmonic mean of the two. Results for individual corpora can be seen in Table~\ref{tab:results_SCD}. 

Note that although we created separate development sets for the two DIHARD corpora, we have found them to be non-representative of the test data -- the optimal thresholds between the development and evaluation sets are \emph{very} different. This is likely due to the rather significant differences between the original challenge's development and evaluation data -- some audio domains were only used for evaluation and some only for development.

To avoid tuning the number of epochs and the decision threshold on test data, we used identical settings for all our models and corpora instead. We set these values in such a way as to obtain high F1 scores on the AMI development set across all four models which were trained or evaluated on AMI -- as five training epochs and a threshold of 0.35.

\begin{figure}[t]
     \centering
 \adjincludegraphics[scale=0.6,trim={0 {0.04\height} 0 {0.03\height}},clip]{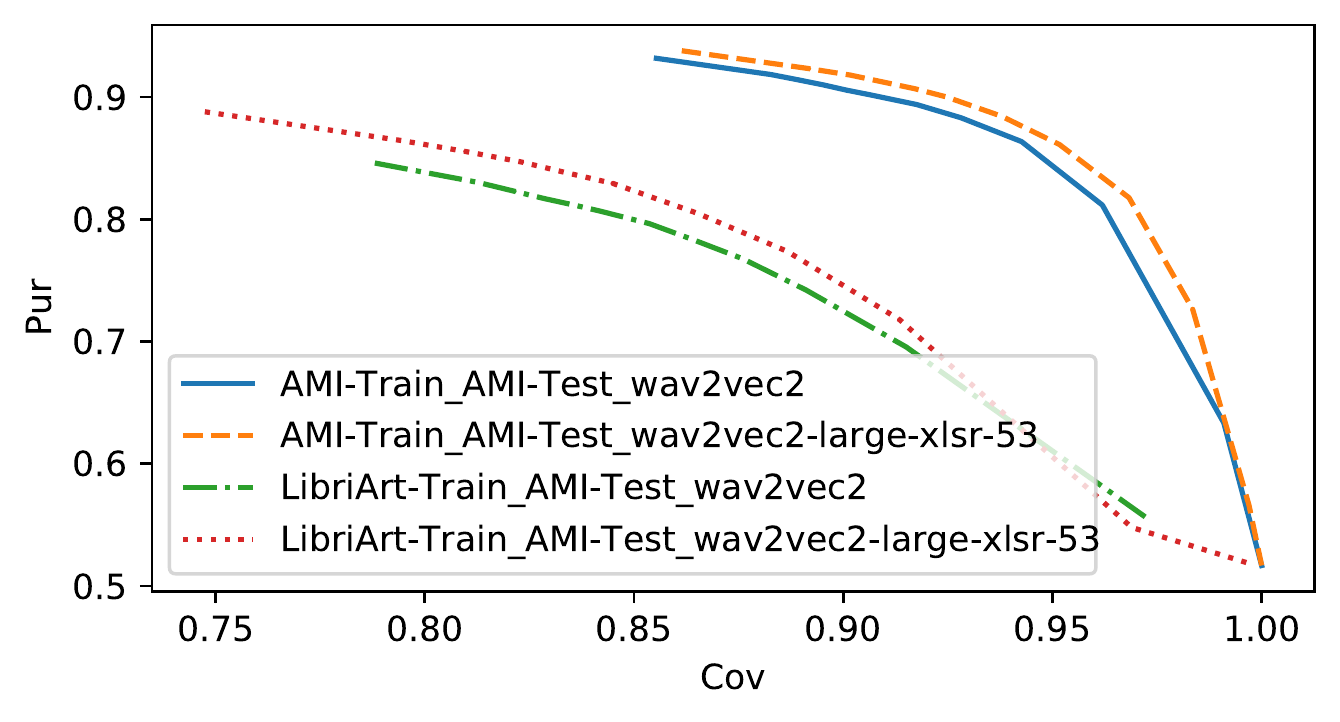}
 \caption{Coverage vs.\@ Purity on the SCD task for the AMI corpus, with model trained on in-domain (AMI) or artificial (``LibriArt'') data.}
 \label{fig:CovPur}
 \end{figure}

The consistency of our tested models is evident from the Coverage vs.\@ Purity graph in Figure~\ref{fig:CovPur} for the AMI corpus.
 
For a comparison with other systems from different state-of-the-art articles, we present Table~\ref{tab:results_SCD_others}, showing the best results on the selected corpora we could find in the literature. 
The results of our models trained on in-domain data all surpass those of the other authors. With the exception of the AMI corpus, models trained on artificial data also achieve comparable results.

\section{Multitask Learning}

After the success of our wav2vec2
approach on the SCD task, we can proceed to the primary goal of this work: extending the basic idea to multitask classification by combining three different
tasks in a single model: SCD, OSD, and VAD. Our expectation here is that the three tasks will complement each other, as they are thematically similar (e.g.,\@ VAD distinguishes between silence and one-or-more speakers, and OSD distinguishes between zero-or-one speakers and multiple). 

Based on our previous findings on the SCD task, we decided to focus our experiments more narrowly: Firstly, for the time being, we limit ourselves to training and testing only on the AMI corpus, as it appears to have the best annotations of the four real corpora, particularly in regard to overlapping speech, and its development set is representative of the test set. We also do not use the artificial training data from section~\ref{subsec_artData}, as they were explicitly tailored for SCD and may not be suitable for training the other two tasks.
Secondly, we only use the base wav2vec2 model. While the larger XLSR model showed a mild improvement on the SCD task, it came at the cost of substantially higher computational requirements, in terms of both memory and time. At the moment, we do not consider the trade-off to be favorable enough to pursue further.

\subsection{Multitask model}

The basic principle of the multitask approach is the same as before, with the main difference that each audio frame is given three independent labels, representing each of the tasks. The final decision layer of the model thus also has three outputs, as seen in Figure~\ref{fig:w2v_diagramZZ}.

Similarly to SCD, the labels for VAD and OSD use a fuzzy labeling function with values between 0 and 1: 1 indicates speech/overlap, 0 indicates non-speech/non-overlap, and there is a linear slope (width 0.4\,s) at the boundaries between the two, as in \cite{Kunesova2019_SPECOM_Springer}. An example of this can be seen in the top two plots of Figure~\ref{fig:multitask_example}. As with SCD, the labels are derived from the RTTM-formatted annotations.

We have also explored two possible changes to the model. Firstly, we tried applying a sigmoid function to the output of the final linear layer. Secondly, during training, we tried to assign different weights to the MSE losses of individual tasks, based on the relative amounts of positive and negative samples in the training data. Unfortunately, neither of these experimental changes had a substantial effect on the performance of our system, so we did not use them in the final version.

\section{Results of the multitask approach}

The results of the multitask model on the AMI corpus are presented in Table~\ref{tab:results_multitask_with_others}. For VAD, the main evaluation metric was the detection error rate (Err), which is calculated as the sum of the miss rate (miss) and false alarm rate (FA), both expressed relative to the total number of \emph{positive} frames (i.e., ground-truth speech or overlap). OSD was also evaluated in terms of precision (Prec) and recall (Rec), F1, and accuracy (Acc), as is common. As with SCD, we do not perform any post-processing of the wav2vec2 predictions -- we only apply a threshold and evaluate the detected intervals with the default \texttt{pyannote.metrics} settings.

The number of training epochs was again five, and each task's threshold was chosen separately based on the highest F1 (OSD, SCD) or lowest detection error rate (VAD) on the development set. 

\input{table_multitask}

While analyzing the outputs of the multitask model, we observed that mismatch between the predicted SCD outputs and the training references is more common in regions with overlapped speech compared to other parts of the audio (also seen on the left side of  Figure~\ref{fig:multitask_example}). The same is true even for the single-task model. This suggests that the model may have trouble detecting speaker changes in the presence of multiple active speakers. However, it is also possible that the reference labels for these regions are simply less accurate. 

In comparison to the single-task model, multitask SCD achieved a slight improvement in F1 score. It is hard to say if this difference is meaningful -- it falls within the range of random fluctuations that we have observed between different training seeds and epochs in the same model. However, it is clear that the system's performance has not suffered from the increased amount of information being represented by a model of the same size. 

For comparison, Table~\ref{tab:results_multitask_with_others} also shows some previous results on the OSD and VAD tasks on the AMI corpus (past results for SCD were shown in Table~\ref{tab:results_SCD_others}). Our multitask model outperforms most of these other systems on the two tasks while remaining the best on SCD. 

\section{Conclusion}


In this paper, we proposed a multi-task approach for three different speech tasks (detection of speaker changes, overlapped speech, and voice activity) using a fine-tuned wav2vec2 model with an additional decision layer. The results of our system on the VAD and OSD tasks are similar to or better than those in the listed previous works, while our SCD results surpass all previous results on the same datasets. Wav2vec2 is a relatively complex model with a high computation cost, but we want to use this approach in a transcription system in combination with existing ASR~\cite{IS22_ASR_wav2vec2_Lehecka}, where the first wav2vec2 layers can be shared. Wav2vec2 can capture high-level information trained unsupervised on huge amounts of acoustic data, which can be used beneficially for our tasks. For future work, our ultimate goal is to directly incorporate the ASR into our multitask approach and train everything together. We expect that the SCD, VAD and OSD will help ASR in multi-speaker scenarios.

\FloatBarrier 

\bibliographystyle{IEEEbib-abbrev} 

\bibliography{bibliography}

\end{document}

%% file: table_SCD_results.tex
\begin{table*}[ht]
\centering


\caption{Our results (\%) for single-task SCD on different corpora, 
with models trained either on in-domain data or on an artificial dataset.} 

\label{tab:results_SCD}

\centering
\small
  
  \begin{tabular}{l *{3}{S[round-mode=places,table-format=2.2, round-precision=2,detect-weight]} *{3}{S[round-mode=places,table-format=2.2, round-precision=2,detect-weight]}}
    \toprule

    \textbf{SCD -- Evaluated corpus} &  \multicolumn{3}{c}{\textbf{Trained on in-domain data}} & \multicolumn{3}{c}{\textbf{Trained on artificial data}}\\
    \cmidrule(lr){2-4}
    \cmidrule(lr){5-7}
    \textbf{and feature model} & {Cov} & {Pur} & {F1} & {Cov} & {Pur} & {F1} \\
                                         
    \midrule 
    
     AMI wav2vec2-base & 
       90.4987 & 90.3654 & 90.432 & 83.1913 & 81.571 & 82.3732 \\ 
     AMI wav2vec2-large-xlsr-53 &
       91.9335 & 90.5934 & 91.2585 & 83.0592 & 84.1016 & 83.5772 \\ 
        \midrule
     DH-I wav2vec2-base & 
     93.4824 & 89.8134 & 91.6112 & 92.7913 & 86.1842 & 89.3658 \\ 
     
     DH-I wav2vec2-large-xlsr-53 & 
     95.2951 & 89.1691 & 92.1304 & 89.1206 & 89.9683 & 89.5424 \\
        \midrule
     DH-II wav2vec2-base & 
     92.5551 & 92.3099 & 92.4323 & 94.8499 & 85.995 & 90.2057 \\
     DH-II wav2vec2-large-xlsr-53 & 
     95.2489 & 91.4102 & 93.2901 & 92.1494 & 89.6883 & 90.9022 \\
    \midrule
     CallHome wav2vec2-base   & 
     93.1465 & 92.9428 & 93.0445 & 92.5294  & 86.5932  & 89.4629 \\ 
     CallHome wav2vec2-large-xlsr-53  & 
     92.9847 & 93.8188 & 93.3999 & 93.4057 & 88.7379 & 91.012 \\ 

    \bottomrule
  \end{tabular}

\end{table*}

%% file: table_SCD_others_results.tex
\begin{table}[t]
\centering

\caption{Previously reported results (\%) for the SCD task on different corpora, with models trained on in-domain data.
}
\label{tab:results_SCD_others}

\centering
\small
  
  \begin{tabular*}{\linewidth}{l l l l} 
    \toprule

    \textbf{SCD -- Evaluated corpus and method} &  {Cov} & {Pur} & {F1} \\
                                         
    \midrule 
     AMI Triplet+content~\cite{ICASSP22_SCD_AMI} & 91.75 & 85.68 & 88.61 \\ 
     AMI Sequence Transd.~\cite{SPL22_SCD_AMI_DH1} & 89.81 & 83.92 & 86.76\\ 
     AMI \texttt{pyannote}~\cite{ICASSP20_Bredin_Pyannote.audio}  & 84.2 & 90.4 & {--}\\ 
    \midrule
     DH-I Sequence Transd.~\cite{SPL22_SCD_AMI_DH1}  & 92.56 & 86.24 & 89.29   \\ 
     \midrule
     DH-II \texttt{pyannote}~\cite{ICASSP20_Bredin_Pyannote.audio}  & 93.7 & 86.8  & {--}\\ 
    \midrule
    CallHome LSTM~\cite{Specom2018_HruzHlavac_SCD_LSTM_Springer} & 72.57 & 72.57 & {--}\\ 
    \bottomrule
    
  \end{tabular*}

\vspace{-0.25em}
\end{table}

%% file: table_multitask.tex
\begin{table}[t]
\centering


\setlength{\tabcolsep}{4pt}

\caption{Results (\%) of the multitask model on the AMI corpus, compared with selected previous results. Thr = threshold.} 

\label{tab:results_multitask_with_others}

\centering
\small
  
\begin{tabularx}{\columnwidth}{X l l l l l l} 


\cmidrule{1-5}
    \textbf{SCD} & {Thr} & {Cov} & {Pur} & {F1} \\
    \cmidrule{1-5}


    ours (single-task) & 0.35 & 90.50 & 90.37 & 90.43\\
    ours (multitask) & 0.40 & 91.68 & 89.91 & 90.79\\


\cmidrule{1-5}

\\ 


\toprule
    \textbf{OSD} & {Thr} & {Prec} & {Rec} & {F1} & {Acc} & {Err}\\
    \midrule

    ours (multitask) & 0.20 & 79.04 & 79.38 & 79.21 & 94.16 & 41.67 \\ 

    \midrule 

    
        \multicolumn{2}{l}{CNN~\cite{Kunesova2019_SPECOM_Springer} (updated in \cite{ICASSP20_Bullock_OSD_AMI_DH1})} & 75.8 & 44.6 & {--} & {--} & {--}\\ 

    
    \multicolumn{2}{l}{\texttt{pyannote} 
    \cite{IS21_OSD_Bredin_AMI_DH3}} & 80.7 & 70.5 & 75.3 & {--} & {--}\\
    \multicolumn{2}{l}{TCN \cite{Cornell2020}} & {--} & {--} & {--} & 95.4 & {--} \\
    
    \multicolumn{2}{l}{x-vectors \cite{Mateju22_interspeech}} &  85.5 & 47.4 & 61.0 & {--} & {--}\\ 
\bottomrule

\\ 


\cmidrule{1-6}
    \textbf{VAD} & {Thr} & {Err} & {miss} & {FA} & {Acc}\\
    \cmidrule{1-6}
    
    ours (multitask) & 0.50 & 5.04 & 1.60 & 3.44 & 95.89\\ 


    \cmidrule{1-6}
    
    \multicolumn{2}{l}{\texttt{pyannote} 
    \cite{IS21_OSD_Bredin_AMI_DH3}} & 6.8 & 3.2 & 3.6 & {--}\\ 
    \multicolumn{2}{l}{TCN~\cite{Cornell2020}} & {--} & {--} & {--} & 94.2 \\
    \multicolumn{2}{l}{wav2vec2~\cite{IS2022_MultiTask_ASR_SD}} &  {--} & 0.7 & 3.1 & {--}  \\ 

\cmidrule{1-6}

\end{tabularx}
\vspace{-0.5em}
  
\end{table}